\documentclass[conference]{IEEEtran}
\IEEEoverridecommandlockouts
\usepackage{epsfig,epstopdf,setspace,subfigure}
\usepackage{amsmath,amssymb,cite,enumerate}
\newtheorem{proposition}{Proposition}

\newtheorem{theorem}{Theorem}

\begin{document}
\title{Alternating-Offer Bargaining Games over the Gaussian Interference Channel}
\author{\IEEEauthorblockN{Xi Liu and Elza Erkip}
\IEEEauthorblockA{ECE Department, Polytechnic Institute of NYU, Brooklyn, NY 11201\\
Email: xliu02@students.poly.edu, elza@poly.edu}
\thanks{This material is based upon work partially supported by NSF Grant No.
0635177, by the Center for Advanced Technology in Telecommunications
(CATT) of Polytechnic Institute of NYU.}
}
\maketitle

\begin{abstract}
This paper tackles the problem of how two selfish users jointly determine the operating point in the achievable rate region of a two-user Gaussian interference channel through bargaining. In previous work, incentive conditions for two users to cooperate using a simple version of Han-Kobayashi scheme was studied and the Nash bargaining solution (NBS) was used to obtain a fair operating point. Here a noncooperative bargaining game of alternating offers is adopted to model the bargaining process and rates resulting from the equilibrium outcome are analyzed. In particular, it is shown that the operating point resulting from the formulated bargaining game depends on the cost of delay in bargaining and how bargaining proceeds. If the associated bargaining problem is regular, a unique perfect equilibrium exists and lies on the individual rational efficient frontier of the achievable rate region. Besides, the equilibrium outcome approaches the NBS if the bargaining costs of both users are negligible.
\end{abstract}

\section{Introduction}

The two-user interference channel (IC) is a fundamental model in information theory for studying interference in communication systems. In this model, each user's transmitter sends an independent message to its corresponding receiver via a common channel and the two communication links interfere with each other. The capacity region for the Gaussian IC is not known in general, but a simplified version of a scheme \cite{references:Etkin08} due to Han and Kobayashi \cite{references:Han81} has recently been shown to result in an achievable rate region that is within one bit of the capacity region for all ranges of channel parameters. However, any type of Han-Kobayashi (H-K) scheme requires full cooperation\footnote{Throughout the paper, ``cooperation'' means cooperation for the choice of transmission strategy including codebook and rate selection and time sharing, which is different from cooperation in information transmission as in cooperative communications\cite{references:Sendonaris03}.} between the two users through the choice of transmission strategy. In practice, users may be selfish in the sense that they choose a transmission strategy to maximize their own rates only. In this case, they may not have an incentive to comply with a certain rule as in the H-K scheme and therefore not all rate pairs in an achievable rate region are actually attainable. Such a scenario becomes increasingly relevant as dynamic spectrum access and sharing becomes important due to the deregulation of wireless spectrum. When there is no coordination among the users, interference is usually treated as noise, which is information theoretically suboptimal in most cases.

When users have conflicting interests, the question of how users interact with each other to achieve efficiency and fairness is usually investigated using game theory. The Gaussian IC was studied using noncooperative game theory in \cite{references:Yu00}\cite{references:Etkin07}\cite{references:Larsson08}, where it was assumed that the receivers treat the interference as Gaussian noise. For the related Gaussian multiple-access channel (MAC), it was shown in \cite{references:Gajic08} that in a noncooperative rate game with two selfish users choosing their transmission rates independently, all points on the dominant face of the capacity region are pure strategy Nash Equilibria (NE). However, no single NE is superior to the others, making it impossible to single out one particular NE to operate at. Noncooperative information theoretical games were considered by Berry and Tse assuming that each user can select any encoding and decoding strategy to maximize its own rate and a Nash equilibrium region was characterized for a class of deterministic IC's \cite{references:Berry08}. Extensions were made to a symmetric Gaussian IC in \cite{references:Berry09}.

Another game theoretical approach for studying interfering links is using the Nash bargaining solution (NBS) from cooperative game theory, e.g., \cite{references:Han05},\cite{references:Mathur_Sankar_Mandayam06}, \cite{references:Leshem08},\cite{references:LiuErkip10}. The advantage of the NBS is that it not only provides a Pareto optimal operating point from the point of view of the entire system, but is also consistent with the fairness axioms of game theory. However, one of the assumptions upon which cooperative bargaining is built is that the users are committed to the agreement reached in bargaining \cite{references:Binmore98}, which requires some form of centralized coordination to ensure that all the parties involved operate at the agreed upon point. In an unregulated environment, a centralized authority may be lacking and in such cases more realistic bargaining between users through communication over a side channel may become necessary. Besides, in most works that designate the NBS as a desired solution, each user's cost of delay in bargaining is not taken into account and little is known regarding how bargaining proceeds. Motivated by all these, we will study the Gaussian IC bargaining problem by introducing a noncooperative bargaining model named the \emph{alternating-offer bargaining game} (AOBG) \cite{references:Myerson91}\cite{references:Martin} from bargaining theory. This approach is different from the NBS in that it models the bargaining process between users explicitly as a non-cooperative multi-stage game in which the users alternate making offers until one is accepted. The equilibrium of such a game describes what bargaining strategies would be adopted by the users and thus provides a nice prediction to the result of noncooperative bargaining. To the best of our knowledge, our approach provides the first application of dynamic AOBG from bargaining theory to network information theory.

Similar to \cite{references:LiuErkip10}, we assume coordination between users is done in two phases. In phase 1, the two users negotiate and only if certain incentive conditions are satisfied they agree to use a particular transmission scheme that can achieve higher rates for both users than treating interference as noise. Such a scheme can be either a simple H-K type scheme from \cite{references:Etkin08} or an orthogonal scheme like TDM or FDM. Conditions under which users can benefit from cooperating using the H-K scheme and the orthogonal one have been investigated in \cite{references:LiuErkip10} and \cite{references:Leshem08} respectively. In phase 2, provided that negotiation in phase 1 is successful, the users bargain over the achievable rate region of the chosen scheme to find an acceptable operating point. This paper differs from \cite{references:LiuErkip10} primarily in that we adopt the AOBG formulation for bargaining instead of the NBS in phase 2. The two-user bargaining problem is considered in an uncoordinated environment where the ongoing bargaining may be interrupted, for example, by other users wishing to access the channel. Each user's cost of delay in bargaining is derived from an exogenous probability which characterizes the risk of breakdown of bargaining due to some outside intervention. The AOBG with risk of breakdown is introduced to model the bargaining process and the negotiation outcome in terms of achievable rates is analyzed. We show that the equilibrium outcome of the AOBG lies on the individual rational efficient frontier of the rate region with its exact location depending on the exogenous probabilities of breakdown. When the breakdown probabilities are very small, it is shown that the equilibrium outcome approaches the Nash solution.

The remainder of this paper is organized as follows. In Section II, we present the channel model, describe the achievable rate regions of a simple H-K type scheme and the TDM scheme, review the concept of AOBG from game theory. We first illustrate how the two users play an AOBG to determine an operating point over the Gaussian MAC in Section III. Then in Section IV we apply the AOBG framework to the Gaussian IC and characterize the equilibrium of the AOBG when the associated bargaining problem is regular. Numerical results are illustrated in Section V. Finally we draw conclusions in Section VI.

\section{System Model}
\subsection{Channel Model}
\begin{figure}
\centering
\includegraphics[width = 2in]{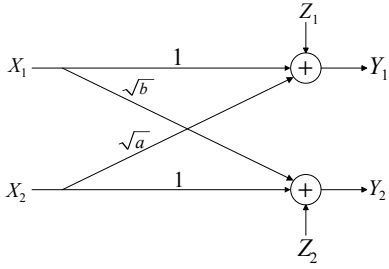}
\caption{Gaussian interference channel}
\label{fig:interference}
\end{figure}

In this paper, we focus on the two-user standard Gaussian IC as shown in Fig. \ref{fig:interference}
\begin{eqnarray}
Y_1 = X_1 + \sqrt{a}X_2+Z_1\\
Y_2 = \sqrt{b}X_1 + X_2 + Z_2
\end{eqnarray}
where $X_i$ and $Y_i$ represent the input and output of user $i \in \{1,2\}$, respectively, and $Z_1$ and $Z_2$ are i.i.d. Gaussian with zero mean and unit variance. Receiver $i$ is only interested in the message sent by transmitter $i$. Constants $\sqrt{a}$ and $\sqrt{b}$ represent the real-valued channel gains of the interfering links. If $a \geq 1$ and $b \geq 1$, the channel is {\em strong} Gaussian IC; if either $0<a<1$ and $b\geq 1$, or $0<b<1$ and $a\geq 1$, the channel is {\em mixed} Gaussian IC; if $0<a<1$ and $0<b<1$, the channel is {\em weak} Gaussian IC. We assume that transmitter of user $i$, $i \in \{1,2\}$, is subject to an average power constraint $P_i$. We let $\text{SNR}_i = P_i$ be the signal to noise ratio (SNR) of user $i$.
\subsection{Achievable Rate Regions}
The best known inner bound for the two-user Gaussian IC is the full H-K achievable region \cite{references:Han81}. Even when the input distributions in the H-K scheme are restricted to be Gaussian, computation of the full H-K region remains difficult due to numerous degrees of freedom involved in the problem \cite{references:Khandani09}. Therefore for the purpose of evaluating and computing bargaining solutions, we assume users employ Gaussian codebooks with equal length codewords and consider a simplified H-K type scheme with fixed power split and no time-sharing as in \cite{references:Etkin08}. Let $\alpha \in [0,1]$ and $\beta \in [0,1]$ denote the fractions of power allocated to the private messages (messages only to be decoded at intended receivers) of user 1 and user 2 respectively. We define $\mathcal{F}$ as the collection of all rate pairs $(R_1,R_2)\in \mathbb{R}^2_{+}$ satisfying
\begin{equation}
R_1 \leq \phi_1 = C\left(\frac{P_1}{1+a\beta P_2}\right)\label{eqn:reg1}
\end{equation}
\begin{equation}
R_2 \leq \phi_2 = C\left(\frac{P_2}{1+b\alpha P_1}\right)\label{eqn:reg2}
\end{equation}
\begin{equation}
R_1 + R_2 \leq \phi_3 = \min\{\phi_{31},\phi_{32},\phi_{33}\}\label{eqn:reg3}
\end{equation}
with
\begin{equation*}
\phi_{31} = C\left(\frac{P_1+a(1-\beta)P_2}{1+a\beta P_2}\right) + C\left(\frac{\beta P_2}{1+b\alpha P_1}\right)
\end{equation*}
\begin{equation*}
\phi_{32} = C\left(\frac{\alpha P_1}{1+a\beta P_2}\right) + C\left(\frac{P_2+b(1-\alpha)P_1}{1+b\alpha P_1}\right)
\end{equation*}
\begin{equation*}
\phi_{33} = C\left(\frac{\alpha P_1+a(1-\beta)P_2}{1+a\beta P_2}\right) + C\left(\frac{\beta P_2+b(1-\alpha)P_1}{1+b\alpha P_1}\right)
\end{equation*}
and
\begin{equation}
\begin{array}{l l}
2R_1+R_2\leq\phi_4&=\displaystyle C\left(\frac{P_1+a(1-\beta)P_2}{1+a\beta P_2}\right) + C\left(\frac{\alpha P_1}{1+a\beta P_2}\right)\\
&\displaystyle +C\left(\frac{\beta P_2+b(1-\alpha)P_1}{1+b\alpha P_1}\right)
\end{array}\label{eqn:reg4}
\end{equation}
\begin{equation}
\begin{array}{l l}
R_1+2R_2\leq\phi_5&=\displaystyle C\left(\frac{P_2+b(1-\alpha)P_1}{1+b\alpha P_1}\right) + C\left(\frac{\beta P_2}{1+b\alpha P_1}\right)\\
&\displaystyle +C\left(\frac{\alpha P_1+a(1-\beta)P_2}{1+a\beta P_2}\right)
\end{array}\label{eqn:reg5}
\end{equation}
where $C(x) = 1/2 \log_2(1+x)$.
The region $\mathcal{F}$ is a polytope and a function of $\alpha$ and $\beta$. We denote the H-K scheme that achieves the rate region $\mathcal{F}$ by $\text{HK}(\alpha,\beta)$. For convenience, we also represent $\mathcal{F}$ in a matrix form as $\mathcal{F} = \{\mathbf{R}|\mathbf{R} \geq \mathbf{0},\: \mathbf{R} \leq \mathbf{R}^1,\: \text{and}\:\mathbf{A}\mathbf{R} \leq \mathbf{B}\}$, where $\mathbf{R} = (R_1\:R_2)^t$, $\mathbf{R}^1 = (\phi_1\:\phi_2)^t$, $\mathbf{B} = (\phi_3\:\phi_4\:\phi_5)^t$, and
\begin{equation}
\mathbf{A} = \left(
\begin{array}{c c c}
1 & 2 & 1\\
 1 & 1 & 2
\end{array}
\right)^t
\end{equation}
Throughout the paper, for any two vectors $\mathbf{U}$ and $\mathbf{V}$, we denote $\mathbf{U} \geq \mathbf{V}$ if and only if $U_i \geq V_i$ for all $i$. $\mathbf{U} \leq \mathbf{V}$, $\mathbf{U} > \mathbf{V}$ and $\mathbf{U} <\mathbf{V}$ are defined similarly.

In the strong interference case, the capacity region is known \cite{refereces:Sato81}\cite{references:Han81} and is achieved by $\text{HK}(0,0)$, i.e., both users send common messages only to be decoded at both destinations. It is the collection of all rate pairs $(R_1,R_2)$ satisfying
\begin{align}
&R_1 \leq C(P_1),\nonumber\\
&R_2 \leq C(P_2),\nonumber\\
&R_1 + R_2 \leq \phi_6 = \min\{C(P_1+aP_2), C(bP_1+P_2)\}\label{eqn:cap_strong}
\end{align}
Note that $\phi_6 = \phi_3$ for $\alpha = \beta = 0$.

For the strong interference case, we choose optimal $\alpha=\beta=0$. For the mixed and weak cases, as in \cite{references:LiuErkip10}, we choose $\alpha$ and $\beta$ as the near optimal power split of \cite{references:Etkin08}, which achieves a rate region that is within one bit to the capacity region. That is, for weak interference $a<1$ and $b<1$, we set $\alpha = \min(1/(bP_1),1)$ and $\beta = \min(1/(aP_2),1)$; for mixed interference $a<1$ and $b\geq 1$, we set $\alpha = 0$ and $\beta = \min(1/(aP_2),1)$.

A simple strategy for the two users to cooperate is through time division avoiding interference. In this case, user $i$ transmits a fraction $\rho_i (0\leq \rho_i\leq 1)$ of the time under the constraint $\rho_1 + \rho_2 \leq 1$. For a given vector $\mathbf{\rho} = (\rho_1\;\rho_2)^t$, the rate obtained by user $i$ is given by $R_i(\mathbf{\rho}) = R_i(\rho_i) = \rho_iC(\frac{P_i}{\rho_i})$. Hence, the TDM rate region is given by
\begin{equation}
\mathcal{R}_{\text{TDM}} = \{\mathbf{R}|\mathbf{R} = (R_1(\rho_1)\;R_2(\rho_2))^t,\; \rho_1+\rho_2\leq 1\}
\label{eqn:tdmreg}
\end{equation}
Note that, unlike the simple H-K scheme, the shape of the TDM rate region does not depend on the cross-link channel gains $\sqrt{a}$ and $\sqrt{b}$.
\subsection{Overview of Bargaining Games}

\subsubsection{Definitions}
A two-player bargaining problem consists of a pair $(\mathcal{G},\mathbf{g}^0)$ where $\mathcal{G}$ is a closed convex subset of $\mathbb{R}^2$, $\mathbf{g}^0 = (g_1^0\; g_2^0)^t$
is a vector in $\mathbb{R}^2$, and the set $\mathcal{G}\cap \{\mathbf{g}|\mathbf{g}\geq \mathbf{g}^0\}$ is nonempty
and bounded. Here $\mathcal{G}$ is the set of all possible payoff allocations or agreements that the two players can jointly achieve, and $\mathbf{g}^0 \in \mathcal{G}$ is the payoff allocation that results if players fail to agree. We refer to $\mathcal{G}$ as the \emph{feasible set} and to $\mathbf{g}^0$ as the \emph{disagreement point}. We say the bargaining problem $(\mathcal{G},\mathbf{g}^0)$ is {\em essential} iff there exists at least one
allocation $\mathbf{g}'$ in $\mathcal{G}$ that is strictly better for both players than $\mathbf{g}^0$, i.e., the set
$\mathcal{G} \cap \{\mathbf{g}|\mathbf{g}>\mathbf{g}^0\}$ is nonempty; we say $(\mathcal{G},\mathbf{g}^0)$
is \emph{regular} iff $\mathcal{G}$ is essential and for any payoff allocation $\mathbf{g}$ in $\mathcal{G}$ \cite{references:Myerson91},
\begin{equation}
\text{if } g_1>g_1^0, \text{ then } \exists \check{\mathbf{g}} \in \mathcal{G} \text{ such that } g_1^0\leq \check{g}_1<g_1 \text{ and } \check{g}_2>g_2,\label{eqn:regular1}
\end{equation}
\begin{equation}
\text{if } g_2>g_2^0, \text{ then } \exists \hat{\mathbf{g}} \in \mathcal{G} \text{ such that } g_2^0\leq \hat{g}_2<g_2 \text{ and } \hat{g}_1>g_1,\label{eqn:regular2}
\end{equation}
Here (\ref{eqn:regular1}) and (\ref{eqn:regular2}) state that whenever a player gets strictly higher payoff than in the disagreement point, then there exists another allocation such that the payoff of the player is reduced while the other player's payoff is strictly increased.

An agreement $\mathbf{g}$ is said to be \emph{efficient} iff there is no agreement in the feasible set $\mathcal{G}$ that makes every player strictly better off. We refer to the set of all efficient agreements as the \emph{efficient frontier} of $\mathcal{G}$. In addition, we refer to the efficient frontier of the individual rational feasible set $\mathcal{G} \cap \{\mathbf{g}|\mathbf{g}\geq \mathbf{g}^0\}$ as the \emph{individual rational efficient frontier}. Given that $\mathcal{G}$ is closed and convex, the regularity conditions in (\ref{eqn:regular1}) and (\ref{eqn:regular2}) hold iff the individual rational efficient frontier is strictly monotone, i.e., it contains no horizonal or vertical line segments.

\subsubsection{The Bargaining Game of Alternating Offers}
In the NBS approach for bargaining, most information concerning the bargaining environment and procedure is abstracted away, and each player's cost of delay in bargaining is not taken into account. A dynamic strategic model of bargaining called the \emph{alternating-offer bargaining game}, on the other hand, provides a detailed description of the bargaining process. In the AOBG, two players take turns in making proposals of payoff allocation in $\mathcal{G}$ until one is accepted or negotiation breaks down.

An important issue regarding modeling of the AOBG is about the cost of delay in bargaining, as it is directly related to players' motives to settle in an agreement rather than insist indefinitely on incompatible demands. In the bargaining game considered in this paper, we derive the cost of delay in bargaining from an exogenous risk of breakdown; i.e., after each round, the bargaining process may terminate in disagreement permanently with an exogenous positive probability if the proposal made in that round gets rejected. In a wireless network, this probability could correspond to the event that other users present in the environment intervene and snatch the opportunity of negotiation on transmission strategies between a pair of users. For example, consider an uncoordinated environment when multiple users operate over a common channel. By default each user's receiver only decodes the intended message from its transmitter and treats the other users' signals as noise. However, groups of users are allowed to coordinate their transmission strategies to improve their respective rates. In the case of a two-user group, if one user's proposal gets rejected by the other user in any bargaining round, it is reasonable to assume that it may terminate the bargaining process with a certain probability and turn to a third user for negotiation.

Consider a regular bargaining problem $(\mathcal{G},\mathbf{g}^0)$ and the two players involved play a dynamic noncooperative game to determine an outcome. Let $p_1$ and $p_2$ be the probabilities of breakdown that satisfy $0<p_1<1$ and $0<p_2<1$. These probabilities of breakdown measure players' costs of delay in bargaining and are assumed to be known by both players. The bargaining procedure of this game is as follows. Player 1 and player 2 alternate making an offer in every odd-numbered round and every even-numbered round respectively. An offer made in each round can be any agreement in the feasible set $\mathcal{G}$. Within each round, after the player whose turn it is to offer announces the proposal, the other player can either accept or reject. In any odd-numbered round, if player 2 rejects the offer made by player 1, there is a probability $p_1$ that the bargaining will end in the disagreement $\mathbf{g}^0$. Similarly, in any even-numbered round, if player 1 rejects the offer made by player 2, there is a probability $p_2$ that the bargaining will end in the disagreement $\mathbf{g}^0$. This process begins from round 1 and continues until some offer is accepted or the game ends in disagreement. When an offer is accepted, an agreement is applied and thus the users get the payoffs specified in the accepted offer. Note in the game described above, the two players only get payoffs at a single round in this game, which is the round at which the bargaining ends in either agreement or disagreement.

For this multi-stage bargaining game, a subgame perfect equilibrium (SPE) is a Nash equilibrium of the whole game with the additional property that the equilibrium strategies induce a Nash equilibrium in every subgame as well. A formal description of the bargaining process in the context of an extensive game with perfect information and chance moves \cite{references:Martin} can be found in the journal version of this paper \cite{references:LiuErkip10_jnl} but will be omitted here.

{\theorem For any regular two-player bargaining problem $(\mathcal{G},\mathbf{g}^0)$, the corresponding AOBG described above has a unique SPE. Let $(\bar{\mathbf{g}},\tilde{\mathbf{g}})$ be the unique pair of efficient agreements in $\mathcal{G}$ which satisfy
\begin{equation}
\tilde{g}_1 = (1-p_2)(\bar{g}_1-g^0_1) + g^0_1 \label{eqn:aobg1}
\end{equation}
\begin{equation}
\bar{g}_2 = (1-p_1)(\tilde{g}_2-g^0_2) + g^0_2\label{eqn:aobg2}
\end{equation}
In the SPE, player 1 always proposes an offer $\bar{\mathbf{g}}$ and accepts any offer $\mathbf{g}$ with $g_1 \geq \tilde{g}_1$; user 2 always proposes an offer $\tilde{\mathbf{g}}$ and accepts any offer $\mathbf{g}$ with $g_2 \geq \bar{g}_2$. Therefore, in equilibrium, the game will end in an agreement on $\bar{\mathbf{g}}$ at round 1.
}
\begin{proof}
The proof of this theorem is similar to that of Theorem 8.3 in \cite{references:Myerson91} with the disagreement outcome fixed to $\mathbf{g}^0$ after the breakdown in any round. Regularity of the bargaining problem is essential for the proof of the uniqueness of the SPE.
\end{proof}

In \cite{references:Binmore86}, it is found that as $p_1$ and $p_2$ approach to zero, the equilibrium outcome of the AOBG converges to the NBS. In other words, if there are no external forces to terminate the bargaining process, the equilibrium outcome of the dynamic game approaches the NBS. More discussion will be given on how the probabilities of breakdown $p_1$ and $p_2$ affect the equilibrium outcome of the bargaining game in the later sections.
\section{Bargaining over the Two-User Gaussian MAC}
Before we move to the Gaussian IC, we first consider a Gaussian MAC in which two users send information to one common receiver. This also forms the foundation for the solution of the strong IC. The received signal is given by
\begin{equation}
Y = X_1 + X_2 + Z
\end{equation}
where $X_i$ is the input signal of user $i$ and $Z$ is Gaussian noise with zero mean and unit variance. Each user has an individual average input power constraint $P_i$. The capacity region $\mathcal{C}$ is the set of all rate pairs $(R_1, R_2)$ such that
\begin{eqnarray}
R_i \leq C(P_i), \: i \in \{1,2\}\\
R_1 + R_2 \leq \phi_0 = C(P_1 + P_2)
\end{eqnarray}
If the two users fully cooperate in codebook and rate selection, any point in $\mathcal{C}$ is achievable. When there is no coordination between users, in the worst case, one user's signal can be treated as noise in the decoding of the other user's signal, leading to rate $R_i^0 = C(\frac{P_i}{1+P_{3-i}})$  for user $i$. In \cite{references:Gajic08}, $R_i^0$ is also called user $i$'s ``safe rate''. If the two users are selfish but willing to coordinate for mutual benefits, they may bargain over $\mathcal{C}$ to determine an operating point with $\mathbf{R}^0$ serving as a disagreement point. 

In this section, we apply the AOBG formulation to the two-user MAC and analyze the negotiation results. For the two-user MAC bargaining problem $(\mathcal{C}_0,\mathbf{R}^0)$, the individual rational efficient frontier is simply the dominant face of the capacity region and is strictly monotone, therefore the regularity conditions in Section II always hold. Using Theorem 1, we have the following proposition.
{\proposition For the two-user MAC bargaining problem $(\mathcal{C}_0,\mathbf{R}^0)$, the unique pair of agreements $(\bar{\mathbf{R}}, \tilde{\mathbf{R}})$ in the SPE of the AOBG is given by
\begin{equation}
(\bar{R}_1\;\bar{R}_2\;\tilde{R}_1\;\tilde{R}_2)^t = M^{-1}(-p_2R_1^0\:\:p_1 R_2^0\:\: \phi_0\:\: \phi_0)^t\label{eqn:spemac}
\end{equation}
where
\begin{equation}
M = \begin{pmatrix}
  1-p_2 & 0 & -1 & 0 \\
  0 & 1 & 0 & -(1-p_1) \\
  1 & 1  & 0 & 0  \\
  0 & 0 & 1 & 1
 \end{pmatrix}
\end{equation}
In equilibrium, the game will end in an agreement on $\bar{\mathbf{R}}$ at round 1.
\label{thm:aobgmac}
}
\begin{proof}
From (\ref{eqn:aobg1}) and (\ref{eqn:aobg2}) in Theorem 1, it follows that the unique pair of agreements $(\bar{\mathbf{R}}, \tilde{\mathbf{R}})$ in the SPE must satisfy
\begin{equation}
\tilde{R}_1 = (1-p_2)(\bar{R}_1-R^0_1) + R^0_1 \label{eqn:aobgmac1}
\end{equation}
\begin{equation}
\bar{R}_2 = (1-p_1)(\tilde{R}_2-R^0_2) + R^0_2\label{eqn:aobgmac2}
\end{equation}
In addition, since $\bar{\mathbf{R}}$ and $\tilde{\mathbf{R}}$ need to be efficient agreements, we have
\begin{equation}
\bar{R}_1 + \bar{R}_2 = \phi_0 \label{eqn:aobgmac3}
\end{equation}
\begin{equation}
\tilde{R}_1 + \tilde{R}_2 = \phi_0 \label{eqn:aobgmac4}
\end{equation}
Solving (\ref{eqn:aobgmac1}), (\ref{eqn:aobgmac2}), (\ref{eqn:aobgmac3}) and (\ref{eqn:aobgmac4}), we obtain the unique pair of agreements $(\bar{\mathbf{R}}, \tilde{\mathbf{R}})$ as in the proposition.
\end{proof}

Clearly, if user 2 makes an offer during the first round instead, the equilibrium outcome would be $\tilde{\mathbf{R}}$. It is not hard to see from (\ref{eqn:aobgmac1}), (\ref{eqn:aobgmac2}) that if $p_1 = p_2 = 0$, then we have $\tilde{\mathbf{R}} = \bar{\mathbf{R}}$.

\begin{figure}
\centering
\includegraphics[width = 3in]{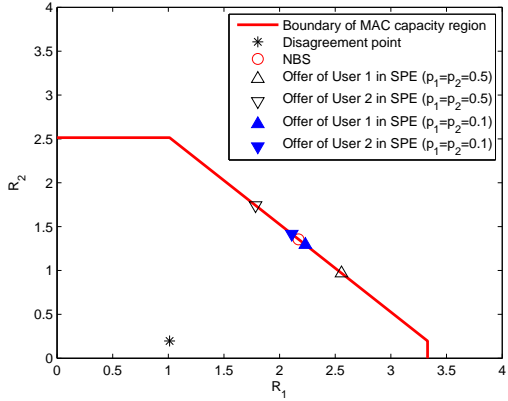}
\caption{Bargaining rates over the MAC when $\text{SNR}_1 = 20$dB, $\text{SNR}_2 = 15$dB}
\label{fig:macnbs}
\end{figure}

In Fig. \ref{fig:macnbs}, the capacity region, the disagreement point and the equilibrium outcomes of the AOBG obtained using Proposition 1 are illustrated for $\text{SNR}_1 = 20$dB and $\text{SNR}_2 = 15$dB. For comparison, the NBS studied in \cite{references:LiuErkip10} is also included in the plot. The unique pairs of agreements $(\bar{\mathbf{R}},\tilde{\mathbf{R}})$ are shown for two different choices of $p_1$ and $p_2$. Recall that offer of user 1 in SPE $\bar{\mathbf{R}}$ corresponds to the equilibrium outcome of the AOBG since we assume user 1 makes an offer first. If user 2 is the first mover instead, offer of user 2 in SPE $\tilde{\mathbf{R}}$ becomes the equilibrium outcome of the game. For a fixed pair of $p_1$ and $p_2$, each user's rate in the equilibrium outcome is higher when it is the first mover than when it is not. Such a phenomenon is referred to as ``first mover advantage'' in \cite{references:Martin}. Finally, as shown in the figure, when $p_1$ and $p_2$ become smaller, both $\tilde{\mathbf{R}}$ and $\bar{\mathbf{R}}$ are closer to the Nash solution.

\begin{figure*}[ht]
\centering
\subfigure[Strong interference]{
\includegraphics[width = 2in]{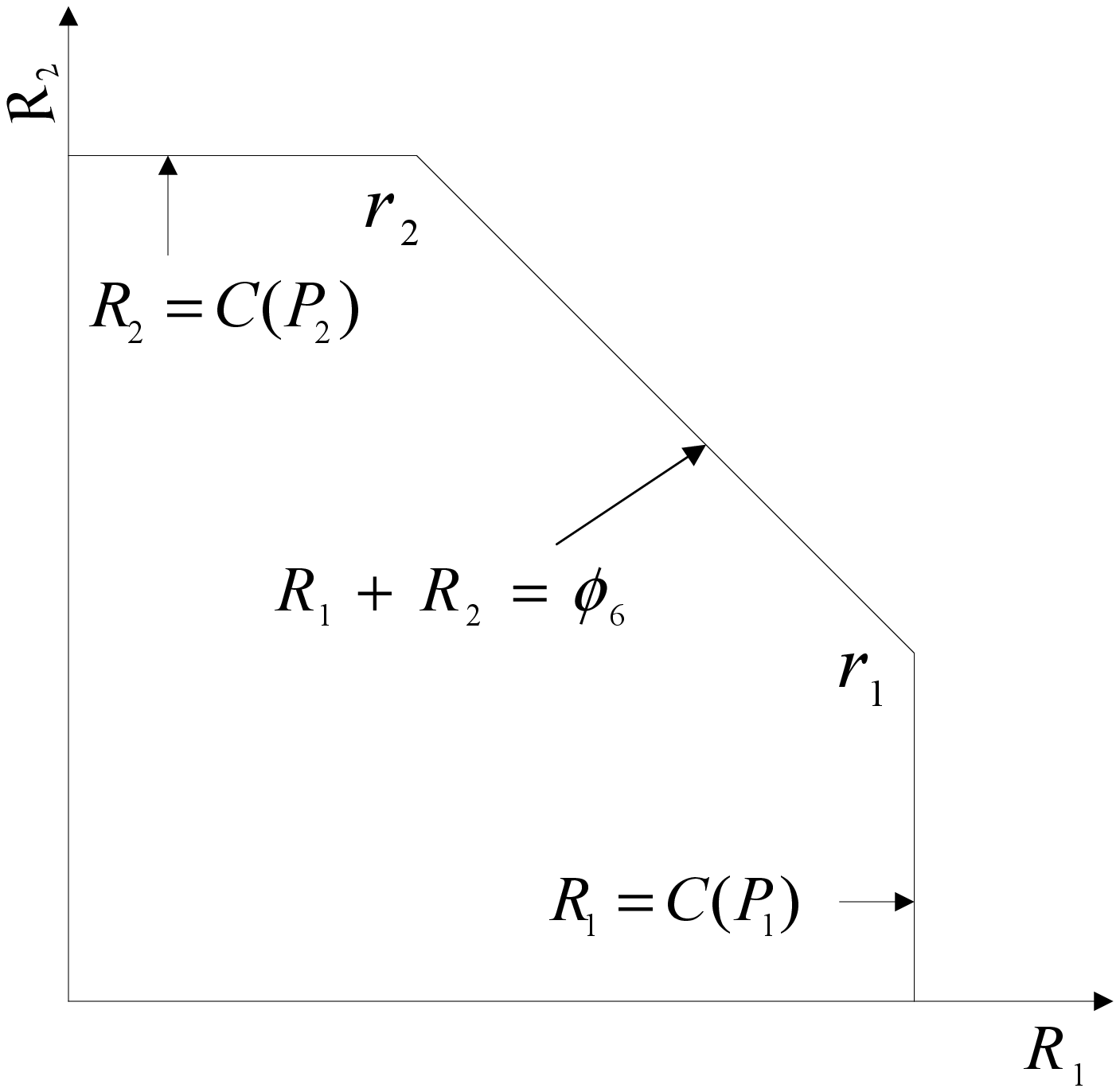}
\label{fig:strreg}
}
\subfigure[Weak or mixed interference]{
\includegraphics[width = 2in]{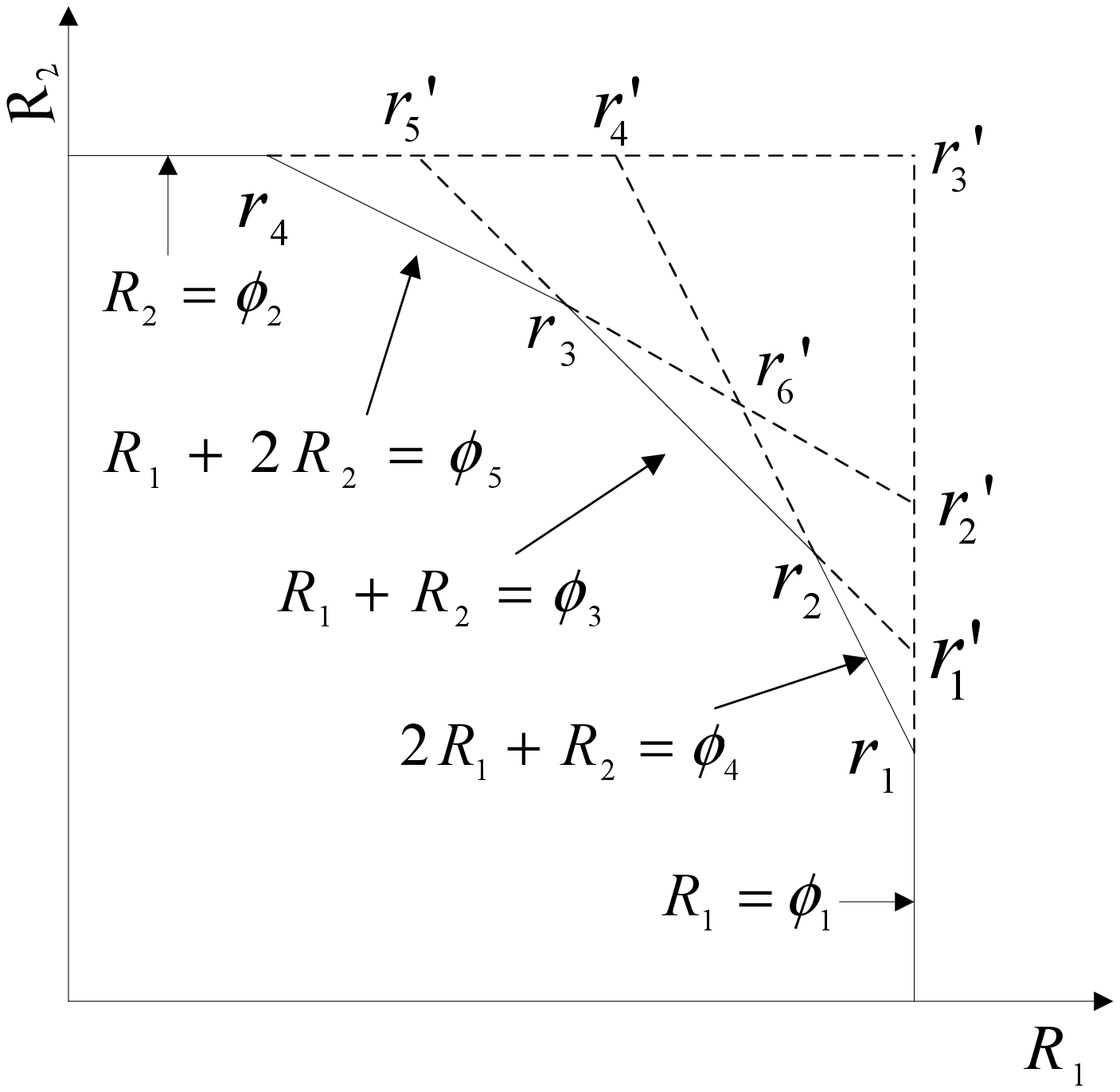}
\label{fig:weakreg}
}
\caption{Achievable rate region using a simple H-K scheme under different interference regimes}
\label{fig:arregions}
\end{figure*}

\section{Two-User Gaussian IC}
For the IC, the coordination between the two users is done in two phases as proposed in \cite{references:LiuErkip10}. In phase 1, users check whether a particular transmission scheme, either a simple H-K type scheme or TDM, improves individual rates for both over those in disagreement $\mathbf{R}^0 = (C(\frac{P_1}{1+aP_2})\; C(\frac{P_2}{1+bP_1}))^t$. If there is no improvement for at least one user, then that user does not have the incentive to cooperate using the chosen scheme and negotiation breaks down. In such a scenario, users operate at the disagreement point $\mathbf{R}^0$; otherwise, they reach an agreement on the use of the chosen scheme and proceed to phase 2. In phase 2, the users bargain for a rate pair to  operate at over the achievable rate region of the scheme they agreed on earlier. If the H-K scheme is employed for cooperation, once a particular rate pair is determined as the bargaining outcome, related codebook information is shared between the users so that one user's receiver can decode the other user's common message as required by the adopted H-K scheme in agreement. If TDM is employed, the time division vector that leads to the rate pair in agreement can be determined accordingly and both users only need to transmit during the portions of time as specified in the vector. No sharing of codebook information is needed in this case. In the following, we discuss the Gaussian IC bargaining problem when the H-K scheme and TDM are employed for cooperation respectively.
\subsection{Cooperating using the H-K scheme}
In this subsection, we assume users employ the simple H-K scheme with optimal or near-optimal power split as discussed in Section II-B for cooperation. The conditions under which both users have incentives to cooperate using this scheme have been studied in \cite{references:LiuErkip10}. We summarize these conditions in the following proposition.

{\proposition For the two-user Gaussian IC, phase 1 is successful and both users have incentives to employ an H-K scheme provided one of the following conditions hold. The conditions also list the H-K scheme employed by the users.
\begin{itemize}
\item Strong interference ($a \geq 1$ and $b \geq 1$): Users always employ HK(0,0);
\item Weak interference ($a <1$ and $b<1$): Users employ HK($1/(bP_1)$,$1/(aP_2)$) iff $aP_2>1$ and $bP_1 > 1$ and $\mathcal{F}\cap \{\mathbf{R}>\mathbf{R}^0\}$ is nonempty when $\alpha = 1/(bP_1)$ and $\beta = 1/(aP_2)$;
\item Mixed interference ($a < 1$ and $b \geq 1$): Users employ HK($0$,$1/(aP_2)$) iff $aP_2>1$ and $\mathcal{F}\cap \{\mathbf{R}>\mathbf{R}^0\}$ is nonempty when $\alpha = 0$ and $\beta = 1/(aP_2)$.
\end{itemize}
\label{thm:incentive}
}
In phase 2, unlike in the MAC case, the associated IC bargaining problem is not always regular. If it is non-regular, the AOBG may have more than one SPE's resulting in distinct bargaining outcomes, which puts any of the SPE's and the corresponding outcome in doubt \cite{references:Martin}. Hence the non-regular case is not treated here. In the following, we first discuss the regularity of the IC bargaining problem under different interference regimes and then characterize the unique SPE of the AOBG when the bargaining problem is regular.
{\proposition Provided that phase 1 is successful, in phase 2, the two-user Gaussian IC bargaining problem $(\mathcal{F}, \mathbf{R}^0)$ is regular iff one of the following conditions hold:
\begin{itemize}
\item Strong interference: $a = b = 1$;
\item Weak interference: $R_1^0 \geq (\phi_5-2\phi_2)^+$ and $R_2^0 \geq (\phi_4-2\phi_1)^+$;

\item Mixed interference: $R_1^0 \geq (\min(\phi_5-2\phi_2,\phi_3-\phi_2))^+$ and $R_2^0 \geq (\min(\phi_4-2\phi_1,\phi_3-\phi_1))^+$;
\end{itemize}
where $\phi_i, i=1,...,5$ are defined in (\ref{eqn:reg1})-(\ref{eqn:reg5}).
\label{thm:regularity}
}
\begin{proof}
In the strong interference case, in phase 1, the users choose optimal $\alpha = \beta = 0$. The resulting capacity region is shown in Fig. \ref{fig:strreg}. Note that only two extreme points of the region are in the first quadrant and they are $r_1 = (\phi_6-C(P_2),C(P_2))$ and $r_2 = (C(P_1),\phi_6-C(P_1))$. It is easy to show that $R_1^0\leq \phi_6-C(P_2)$ and $R_2^0 \leq \phi_6-C(P_1)$ with equalities holding only when $a = b = 1$. In order for the individual rational efficient frontier to be strictly monotone, it must contain no horizonal or vertical line segments, which requires $R_1^0\geq \phi_6-C(P_2)$ and $R_2^0 \geq \phi_6-C(P_1)$. Hence, the associated bargaining problem is regular iff $a = b = 1$.

In the weak interference case, by Proposition \ref{thm:incentive}, in phase 1, both users have incentives to cooperate using $\text{HK}(1/(bP_1),1/(aP_2))$ if $aP_2>1$, $bP_1 > 1$ and $\mathcal{F}\cap \{\mathbf{R}>\mathbf{R}^0\}$ is nonempty when $\alpha = 1/(bP_1)$ and $\beta = 1/(aP_2)$. The shape of achievable rate region is shown in Fig. \ref{fig:weakreg}. It has been proved in \cite{references:Khandani09} that the points $r_i'\notin\mathcal{F}$ for $i\in \{1,2,...,6\}$. Therefore there are at most\footnote{In \cite{references:Khandani09}, the authors concluded that there should be exactly four extreme points in the first quadrant, but we find that under some parameters one or two of the four points may actually not lie in the first quadrant. For instance, it is possible that $\phi_5-2\phi_2<0$, in which case $r_4$ is not in the first quadrant.} four extreme points in the first quadrant of Fig. \ref{fig:weakreg}, given by
\begin{align}
&r_1 = (\phi_1,\phi_4-2\phi_1)\\
&r_2 = (\phi_4-\phi_3,2\phi_3-\phi_4)\\
&r_3 = (2\phi_3-\phi_5,\phi_5-\phi_3)\\
&r_4 = (\phi_5-2\phi_2,\phi_2)
\end{align}
where $\phi_i,\: i\in\{1,2,...,5\}$ are given in (\ref{eqn:reg1})-(\ref{eqn:reg5}) with $\alpha = 1/(bP_1)$ and $\beta = 1/(aP_2)$.
In order for the individual rational efficient frontier to be strictly monotone, it must contain no horizonal or vertical line segments. If $r_1$ is in the first quadrant, $R_2^0\geq \phi_4-2\phi_1$ must hold and similarly if $r_4$ is in the first quadrant, $R_1^0\geq \phi_5-2\phi_2$ must hold.
Hence, the associated bargaining problem is regular iff two additional conditions $R_1^0 \geq (\phi_5-2\phi_2)^+$ and $R_2^0 \geq (\phi_4-2\phi_1)^+$ are satisfied. Here $(\cdot)^+$ means $\max(\cdot,0)$.

In the mixed interference case, by Proposition \ref{thm:incentive}, in phase 1, both users cooperate using $\text{HK}(0,1/(aP_2))$ if $aP_2>1$ and $\mathcal{F}\cap \{\mathbf{R}>\mathbf{R}^0\}$ is nonempty when $\alpha = 0$ and $\beta = 1/(aP_2)$. Similar to the weak interference case, there are at most four extreme points in the first quadrant of Fig. \ref{fig:weakreg} except that $r_1' = (\phi_1,\phi_3-\phi_1)$ or $r_5'= (\phi_3-\phi_2,\phi_2)$ may become an extreme point of $\mathcal{F}$, depending on whether the constraint (\ref{eqn:reg4}) or (\ref{eqn:reg5}) is redundant or not respectively. In order for the individual rational efficient frontier to be strictly monotone, it must contain no horizonal or vertical line segments. If $r_1$ and $r_1'$ are both in the first quadrant, $R_2^0\geq \min(\phi_4-2\phi_1,\phi_3-\phi_1)$ must hold and if $r_4$ and $r_5'$ are both in the first quadrant, $R_1^0\geq \min(\phi_5-2\phi_2,\phi_3-\phi_2)$ must hold. Hence, the associated bargaining problem is regular iff two additional conditions $R_1^0 \geq (\min(\phi_5-2\phi_2,\phi_3-\phi_2))^+$ and $R_2^0 \geq (\min(\phi_4-2\phi_1,\phi_3-\phi_1))^+$ are satisfied.
\end{proof}

When phase 1 is successful and the Gaussian IC bargaining problem $(\mathcal{F},\mathbf{R}^0)$ is regular, using Theorem 1, we have the following result.
{\proposition  For any regular Gaussian IC bargaining problem $(\mathcal{F},\mathbf{R}^0)$, the unique pair of agreements $(\bar{\mathbf{R}}, \tilde{\mathbf{R}})$ and the equilibrium strategies in the SPE of the AOBG are characterized in Theorem 1 with $\mathcal{G} =\mathcal{F}$ of Section II-B, $\mathbf{g}^0 = \mathbf{R}^0 = (C(\frac{P_1}{1+aP_2})\;C(\frac{P_2}{1+bP_1}))^t$, $\bar{\mathbf{g}} = \bar{\mathbf{R}}$ and $\tilde{\mathbf{g}} = \tilde{\mathbf{R}}$.
}

In the strong interference case $a = b = 1$, the unique pair of agreements $(\bar{\mathbf{R}}, \tilde{\mathbf{R}})$ in the SPE can be obtained using (\ref{eqn:spemac}) in Proposition \ref{thm:aobgmac} with $\phi_0$ replaced by $\phi_6$. For the weak and mixed interference cases, since the shape of the H-K rate region and the relative location of the disagreement point vary as parameters $a$, $b$, $P_1$ and $P_2$ change, it is difficult to obtain a general expression for $(\bar{\mathbf{R}}, \tilde{\mathbf{R}})$. However, when all the parameters are given and the corresponding power split parameters $\alpha$ and $\beta$ are fixed, the H-K rate region and the disagreement point $\mathbf{R}^0$ can be determined accordingly. Since $\bar{\mathbf{R}}$ and $\tilde{\mathbf{R}}$ both lie on the individual rational efficient frontier of $\mathcal{F}$ which is piecewise linear, we can compute $(\bar{\mathbf{R}}, \tilde{\mathbf{R}})$ by solving linear equations.

\subsection{Cooperating using TDM}
When the two users cooperate using TDM, the achievable rate region is $\mathcal{R}_{TDM}$ as defined in Section II. In the strong interference case, TDM is strictly suboptimal in terms of achievable rate region; however in the mixed and weak interference cases, the H-K scheme does not always dominate TDM. Due to simplicity of implementation of TDM compared with the H-K scheme, we also investigate the AOBG formulation when TDM is employed for cooperation.

In phase 1, both users have incentives to cooperate using TDM if $(\mathcal{R}_{TDM}, \mathbf{R}^0)$ is essential, i.e., if there exists at least one time division vector $\mathbf{\rho} = (\rho_1\;\rho_2)^t$ such that $R_i(\rho_i)>R_i^0$, for $i = 1,2$. In other words, $\mathbf{R}^0$ must lie strictly inside $\mathcal{R}_{TDM}$. Otherwise, at least one user would not have the incentive to cooperate using TDM and negotiation breaks down. More discussion on conditions under which users can benefit from cooperating using an orthogonal scheme can be found in \cite{references:Leshem08}.  Since the Pareto boundary of $\mathcal{R}_{TDM}$ corresponds to $\rho_1+\rho_2 = 1$ and is strictly monotone, the bargaining problem $(\mathcal{R}_{TDM}, \mathbf{R}^0)$ is regular as long as it is essential.

For the TDM case, using Theorem 1, we have the following results for the AOBG played in phase 2.
{\proposition  For any essential bargaining problem $(\mathcal{R}_{TDM},\mathbf{R}^0)$ over the two-user Gaussian IC, the unique pair of agreements $(\bar{\mathbf{R}}, \tilde{\mathbf{R}})$ and the equilibrium strategies in the SPE of the AOBG are characterized in Theorem 1 with $\mathcal{G} =\mathcal{R}_{TDM}$ in (\ref{eqn:tdmreg}), $\mathbf{g}^0 = \mathbf{R}^0 = (C(\frac{P_1}{1+aP_2})\;C(\frac{P_2}{1+bP_1}))^t$, $\bar{\mathbf{g}} = \bar{\mathbf{R}}$ and $\tilde{\mathbf{g}} = \tilde{\mathbf{R}}$.
}

Unlike the H-K case, the boundary of the TDM rate region is not linear; however, the unique pair of $(\bar{\mathbf{R}}, \tilde{\mathbf{R}})$ can be computed numerically.

\begin{figure}
\centering
\includegraphics[width = 3in]{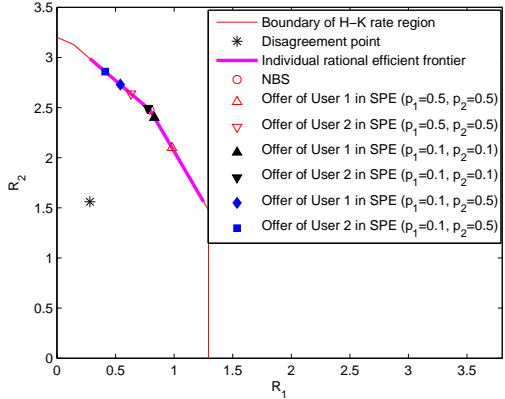}
\caption{The NBS and equilibrium outcomes of AOBG for IC under mixed interference with $a = 0.2$, $b = 1.2$, $\text{SNR}_1 = 10$dB and $\text{SNR}_2 = 20$dB.}
\label{fig:icaobg}
\end{figure}
\begin{figure}
\centering
\includegraphics[width = 3in]{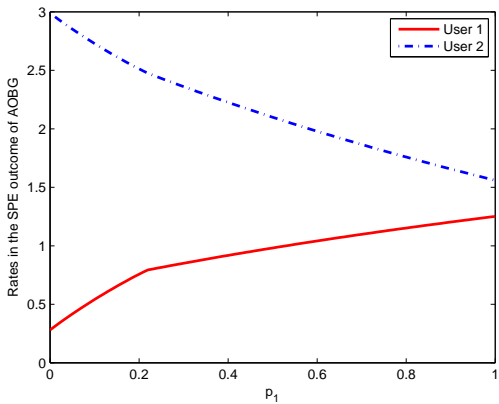}
\caption{Rate of each user in SPE of AOBG as a function of breakdown probability $p_1$ when $p_2 = 0.5$ for IC under mixed interference with $a = 0.2$, $b = 1.2$, $\text{SNR}_1 = 10$dB and $\text{SNR}_2 = 20$dB.}
\label{fig:speratevsp1}
\end{figure}
\section{Illustration of Results}
In Fig. \ref{fig:icaobg}, the unique pair of agreements $(\bar{\mathbf{R}}, \tilde{\mathbf{R}})$ in the SPE of the AOBG based on the H-K scheme is shown for mixed interference with $a = 0.2$, $b = 1.2$, $\text{SNR}_1 = 10$dB and $\text{SNR}_2 = 20$dB for three different choices of the pair of probabilities of breakdown $p_1$ and $p_2$. According to Proposition \ref{thm:incentive}, in phase 1, the two users decide to cooperate using $\text{HK}(0,0.05)$. Furthermore, by Proposition \ref{thm:regularity}, the bargaining problem in phase 2 is regular. As in the MAC case, user 1's offer in SPE $\bar{\mathbf{R}}$ corresponds to the equilibrium outcome of the AOBG since we assume user 1 makes an offer first. If user 2 moves first instead, user 2's offer in SPE $\tilde{\mathbf{R}}$ would become the equilibrium outcome of the game. We can see that as $p_1$ and $p_2$ change, $\bar{\mathbf{R}}$ and $\tilde{\mathbf{R}}$ move along the individual rational efficient frontier of $\mathcal{F}$. When $p_1 = 0.5$ and $p_2 = 0.5$, user 1's rate in $\bar{\mathbf{R}}$ is greater than that in the NBS; but when $p_1 = 0.1$ and $p_2 = 0.5$, its rate in $\bar{\mathbf{R}}$ is smaller than that in the NBS. As both $p_1$ and $p_2$ decrease to $0.1$, both $\bar{\mathbf{R}}$ and $\tilde{\mathbf{R}}$ become closer to the Nash solution. The rate of each user in the equilibrium outcome $\bar{\mathbf{R}}$ as a function of breakdown probability $p_1$ is plotted in Fig. \ref{fig:speratevsp1} when $p_2$ is fixed to 0.5 under the same channel parameters. As $p_1$ gets larger, user 1's rate increases while user 2's decreases. The larger $p_1$ becomes, the more likely that bargaining may terminate in disagreement when user 1's offer is rejected by user 2. This demonstrates that if user 1 fears less about the bargaining breakdown, it can be more advantageous in bargaining. It should also be emphasized that the equilibrium is unique and agreement is reached in round 1 in equilibrium. In this sense, the bargaining mechanism of AOBG is highly efficient.

\begin{figure*}[ht]
\centering
\subfigure[$a = 0.1$, $b = 1.2$]{
\includegraphics[width = 3in]{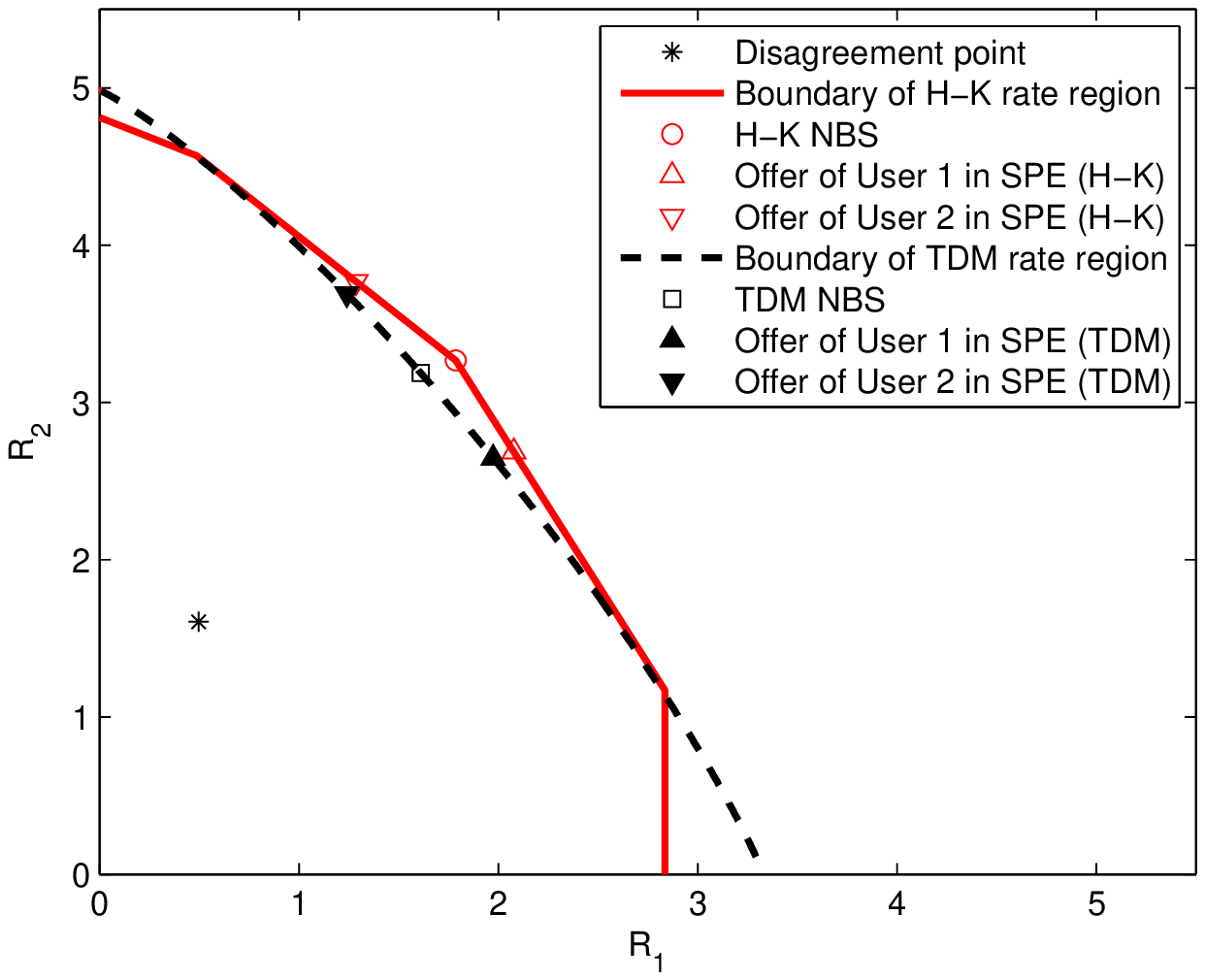}
\label{fig:aobgtdm1}
}
\subfigure[$a = 0.2$, $b = 1.2$]{
\includegraphics[width = 3in]{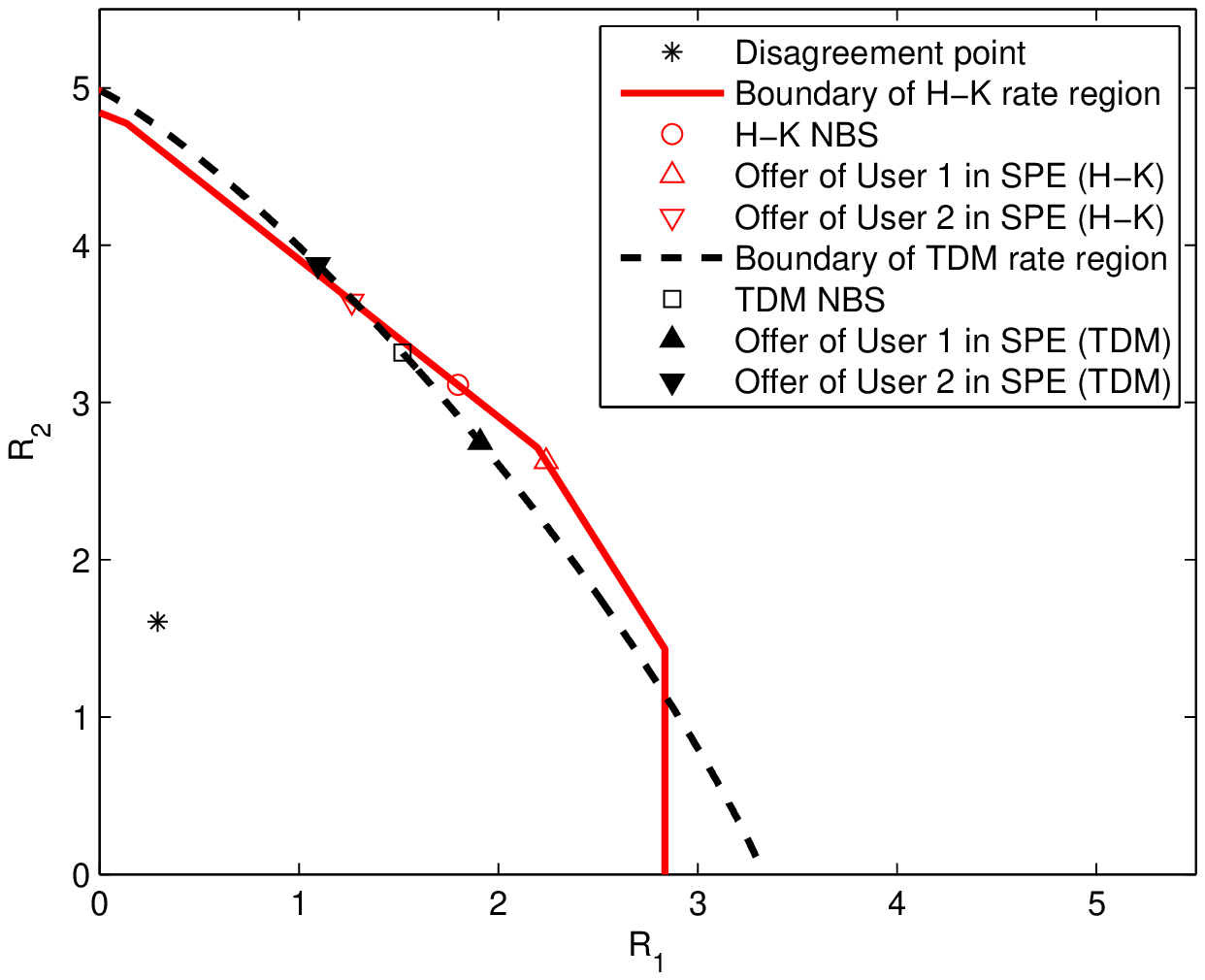}
\label{fig:aobgtdm2}
}
\caption{Comparison of bargaining outcomes when the H-K and TDM schemes are used respectively under mixed interference with $\text{SNR}_1 = 20$dB, $\text{SNR}_2 = 30$dB and $p_1 = p_2 =0.5$.}
\label{fig:aobgtdm}
\end{figure*}
Fig. \ref{fig:aobgtdm} illustrates the equilibrium outcomes of the AOBG when the H-K and TDM cooperating schemes are used respectively for two different channel gain vectors under mixed interference when $\text{SNR}_1 = 20$dB and $\text{SNR}_2 = 30$dB. The probabilities of breakdowns are set as $p_1 = p_2 = 0.5$. The NBS's in both cases are also plotted for reference. In Fig. \ref{fig:aobgtdm1}, $a = 0.1$ and $b = 1.2$, hence in the H-K scheme the power splits are fixed to be $\alpha=0$ and $\beta = 0.01$. In this case, the individual rational efficient frontier in H-K strictly dominates the one in TDM and thus both users' rates in all the bargaining outcomes are superior to those in TDM. In Fig. \ref{fig:aobgtdm2}, the channel parameters are set to $a = 0.2$ and $b = 1.2$ and $\text{HK}(0,0.005)$ is employed. The individual rational efficient frontiers in H-K and TDM intersect. We observe that while user 2 gets higher rates in all the bargaining outcomes in TDM than in H-K, user 1's rates in H-K are superior to those in TDM. Hence, we can conclude that, depending on the channel parameters and power constraints, the two users may have distinct preferences between the transmission schemes employed.

\section{Conclusions}
In this paper, we investigated the problem of how two selfish users over the Gaussian IC coordinate their transmission strategies to boost their rates. The two users first negotiate for the use of either a simple H-K type scheme or TDM in phase 1 and then they bargain over the achievable rate region of the chosen scheme to determine a point to operate at in phase 2. Unlike in the previous work where the NBS is used to select a fair operating point, the dynamic AOBG is adopted to model the bargaining process and determine a bargaining outcome in phase 2. As a problem of independent interest, and also as a tool for developing the optimal solution in the strong interference regime, we first study the MAC before moving on to the IC. The results from the dynamic AOBG show that the bargaining game has a unique SPE and in equilibrium the agreement is reached immediately in the first bargaining round provided that the associated bargaining problem is regular. The exogenous probabilities of breakdown and which user makes a proposal first also play important roles in the final outcome. When the cost of delay in bargaining is not negligible, that is, exogenous probabilities of breakdown are high, the equilibrium outcome deviates from the NBS. We conclude that when we consider coordination and bargaining over the IC, factors such as the cost of delay in bargaining and the environment in which bargaining takes place should also be taken into consideration.

Regarding possible extensions to this work, it would be also interesting to model the cost of delay in bargaining under other assumptions such as each user's payoff is discounted by a factor of $\delta$ after each round \cite{references:Rubinstein82}\cite{references:Binmore86} or the amount of communication overhead incurred. In addition, the bargaining framework here can be extended to the two-user MIMO IC using the results of \cite{references:Vishwanath04}\cite{references:Shang09_archive}.

\bibliographystyle{IEEEtran}
\bibliography{IEEEabrv,references3}

\begin{thebibliography}{10}
\providecommand{\url}[1]{#1}
\csname url@samestyle\endcsname
\providecommand{\newblock}{\relax}
\providecommand{\bibinfo}[2]{#2}
\providecommand{\BIBentrySTDinterwordspacing}{\spaceskip=0pt\relax}
\providecommand{\BIBentryALTinterwordstretchfactor}{4}
\providecommand{\BIBentryALTinterwordspacing}{\spaceskip=\fontdimen2\font plus
\BIBentryALTinterwordstretchfactor\fontdimen3\font minus
  \fontdimen4\font\relax}
\providecommand{\BIBforeignlanguage}[2]{{%
\expandafter\ifx\csname l@#1\endcsname\relax
\typeout{** WARNING: IEEEtran.bst: No hyphenation pattern has been}%
\typeout{** loaded for the language `#1'. Using the pattern for}%
\typeout{** the default language instead.}%
\else
\language=\csname l@#1\endcsname
\fi
#2}}
\providecommand{\BIBdecl}{\relax}
\BIBdecl

\bibitem{references:Etkin08}
R.~Etkin, D.~Tse, and H.~Wang, ``Gaussian interference channel capacity to
  within one bit,'' \emph{{IEEE} Trans. Inf. Theory}, vol.~54, no.~12, pp.
  5534--5562, 2008.

\bibitem{references:Han81}
T.~S. Han and K.~Kobayashi, ``A new achievable rate region for the interference
  channel,'' \emph{{IEEE} Trans. Inf. Theory}, vol.~27, no.~1, pp. 49--60,
  1981.

\bibitem{references:Sendonaris03}
A.~Sendonaris, E.~Erkip, and B.~Aazhang, ``User cooperation diversity - {P}art
  {I}: {S}ystem description,'' \emph{{IEEE} Trans. Commun.}, vol.~51, no.~11,
  pp. 1927--1938, 2003.

\bibitem{references:Yu00}
W.~Yu, G.~Ginis, and J.~Cioffi, ``Distributed multiuser power control for
  digital subscriber lines,'' \emph{{IEEE} J. Sel. Areas Commun.}, vol.~20,
  no.~5, pp. 1105--1115, 2002.

\bibitem{references:Etkin07}
R.~Etkin, A.~Parekh, and D.~Tse, ``Spectrum sharing for unlicensed bands,''
  \emph{{IEEE} J. Sel. Areas Commun.}, vol.~25, no.~3, pp. 517--528, 2007.

\bibitem{references:Larsson08}
E.~G. Larsson and E.~A. Jorswieck, ``Competition versus cooperation on the
  {MISO} interference channel,'' \emph{{IEEE} J. Sel. Areas Commun.}, vol.~26,
  no.~7, pp. 1059--1069, 2008.

\bibitem{references:Gajic08}
V.~Gajic and B.~Rimoldi, ``Game theoretic considerations for the {G}aussian
  multiple access channel,'' in \emph{Proceedings of IEEE ISIT}, Toronto,
  Canada, July 2008, pp. 2523--2527.

\bibitem{references:Berry08}
R.~Berry and D.~Tse, ``Information theoretic games on interference channels,''
  in \emph{Proceedings of IEEE ISIT}, Toronto, Canada, July 2008.

\bibitem{references:Berry09}
------, ``Information theory meets game theory on the interference channel,''
  in \emph{Proceedings of IEEE ITW}, Volos, Greece, June 2009.

\bibitem{references:Han05}
Z.~Han, Z.~Ji, and K.~J.~R. Liu, ``Fair multiuser channel allocation for
  {OFDMA} networks using {N}ash bargaining solutions and coalitions,''
  \emph{{IEEE} Trans. Commun.}, pp. 1366--1376, Aug. 2005.

\bibitem{references:Mathur_Sankar_Mandayam06}
S.~Mathur, L.~Sankar, and N.~B. Mandayam, ``Coalitional games in {G}aussian
  interference channels,'' in \emph{Proceedings of IEEE ISIT}, Jul. 2006.

\bibitem{references:Leshem08}
A.~Leshem and E.~Zehavi, ``Cooperative game theory and the {G}aussian
  interference channel,'' \emph{{IEEE} J. Sel. Areas Commun.}, vol.~26, no.~7,
  pp. 1078--1088, 2008.

\bibitem{references:LiuErkip10}
X.~Liu and E.~Erkip, ``Coordination and bargaining over the {G}aussian
  interference channel,'' in \emph{Proceedings of IEEE ISIT}, Jun. 2010.

\bibitem{references:Binmore98}
K.~G. Binmore, \emph{Game Theory and the Social Contract, Vol 2: Just
  Playing}.\hskip 1em plus 0.5em minus 0.4em\relax The MIT Press, 1998.

\bibitem{references:Myerson91}
R.~B. Myerson, \emph{Game Theory}.\hskip 1em plus 0.5em minus 0.4em\relax
  Harvard University Press, 1991.

\bibitem{references:Martin}
M.~J. Osborne and A.~Rubinstein, \emph{A course in game theory}.\hskip 1em plus
  0.5em minus 0.4em\relax The MIT Press, 1994.

\bibitem{references:Khandani09}
A.~S. Motahari and A.~K. Khandani, ``Capacity bounds for the {G}aussian
  interference channel,'' \emph{{IEEE} Trans. Inf. Theory}, vol.~55, no.~2, pp.
  620--643, 2009.

\bibitem{refereces:Sato81}
H.~Sato, ``The capacity of the {G}aussian interference channel under strong
  interference,'' \emph{{IEEE} Trans. Inf. Theory}, vol.~27, pp. 786--788, Nov.
  1981.

\bibitem{references:LiuErkip10_jnl}
X.~Liu and E.~Erkip, ``A game-theoretic view of the interference channel:
  impact of coordination and bargaining,'' submitted to \emph{IEEE Trans. Inf.
  Theory}, available on line, 2010.

\bibitem{references:Binmore86}
K.~G. Binmore, A.~Rubinstein, and A.~Wolinsky, ``The {N}ash bargaining solution
  in economic modelling,'' \emph{Rand Journal of Economics}, vol.~17, no.~2,
  pp. 176--188, 1986.

\bibitem{references:Rubinstein82}
A.~Rubinstein, ``Perfect equilibrium in a bargaining model,''
  \emph{Econometrica}, vol.~50, pp. 97--109, 1982.

\bibitem{references:Vishwanath04}
S.~Vishwanath and S.~A. Jafar, ``On the capacity of vector {G}aussian
  interference channels,'' in \emph{Proceedings of IEEE ITW}, Oct 2004, pp.
  689--699.

\bibitem{references:Shang09_archive}
\BIBentryALTinterwordspacing
X.~Shang, B.~Chen, G.~Kramer, and H.~V. Poor, ``Capacity regions and sum-rate
  capacities of vector {G}aussian interference channels,'' 2009. [Online].
  Available: \url{http://arxiv.org/abs/0907.0472}
\BIBentrySTDinterwordspacing

\end{thebibliography}
\end{document}